\documentclass[twocolumn,prl,showpacs]{revtex4}
\usepackage{graphicx}
\usepackage{dcolumn}
\usepackage{bm}
\usepackage{amsmath}
\usepackage{times}

\begin{document}

\preprint{}
\title{Angular dependence of Josephson currents in unconventional
superconducting junctions}
\author{T. Yokoyama$^1$, Y. Sawa$^1$, Y. Tanaka$^1$ and A. A. Golubov$^2$ }
\affiliation{$^1$Department of Applied Physics, Nagoya University, Nagoya, 464-8603, Japan%
\\
and CREST, Japan Science and Technology Corporation (JST) Nagoya, 464-8603,
Japan \\
$^2$ Faculty of Science and Technology, University of Twente, 7500 AE,
Enschede, The Netherlands}
\date{\today}

\begin{abstract}
Josephson effect in junctions between unconventional superconductors is
studied theoretically within the model describing the effects of interface
roughness. The particularly important issue of applicability of the
frequently used Sigrist-Rice formula for Josephson current in $d $-wave
superconductor / insulator / $d$-wave superconductor junctions is addressed.
We show that although the SR formula is not applicable in the ballistic
case, it works well for rough interfaces when the diffusive normal metal
regions exist between the $d$-wave superconductor and the insulator. It is
shown that the SR approach only takes into account the component of the $d$%
-wave pair potential symmetric with respect to an inversion around the plane
perpendicular to the interface. Similar formula can be derived for general
unconventional superconductors with arbitrary angular momentum $l$.
\end{abstract}

\maketitle




%

%




A number of phase-sensitive experiments have convincingly demonstrated the
realization of $d$-wave pairing state in high $T_{C}$ cuprates \cite%
{Tsuei,Harlin,SR95,Kashiwaya}. Because of such unconventional symmetry, the
study of Josephson effect in high $T_{C}$ superconducting (HTS) junctions
attracted a lot of interest. A while ago simple formula for the Josephson
current of $d$-wave superconductor / insulator / $d$-wave superconductor
(DID) junctions was proposed by Sigrist and Rice (SR)\cite{Sigrist2}.
According to the SR formula, the Josephson current is proportional to $\cos
2\alpha \cos 2\beta $, where the $\alpha (\beta )$ denotes the angle between
the normal to the interface and the crystal axis of the left(right) $d$-wave
superconductor \cite{Sigrist2}.

Although the SR formula can explain experiments with the so-called $\pi$
-junctions \cite{Tsuei,Harlin}, this formula does not take into account the
effect of mid-gap Andreev resonant states (MARS) formed at junction
interfaces~\cite{Buch,Tanaka95}. Actually, as shown in Ref. \onlinecite{TK97}
, SR formula does not work in ballistic $d$-wave junctions for $\alpha \neq 0$   and $\beta \neq 0$ where MARS influence severely the charge transport at
low temperatures. It was shown both theoretically \cite{TK,Barash} and
experimentally \cite{Ilichev,Testa} that MARS induce a nonmonotonic
temperature dependence of the maximum Josephson current in DID junctions. On
the other hand, SR formula has been extensively used to analyze experiments
with various types of HTS Josephson junctions \cite{Lombardi,Bauch,Lee}.
Experiments with HTS\ junctions are of high importance for basic
understanding of high-Tc superconductivity since they may provide an
information on possible subdominant admixtures to the $d$-wave symmetry\cite%
{Harlingen,Smilde,Tsuei2}. Therefore it is of fundamental interest to
understand the physical mechanisms which determine the angular
dependence of Josephson current in HTS junctions. For this reason, 
the determination of the conditions of applicability of the SR
formula is an important issue which is addressed in the present
paper.

%
%
%
In the following, we study the Josephson current in D/DN/I/DN/D junctions,
where DN denotes diffusive normal metal and could be formed between the insulator and $d$-wave superconductors.
The calculations are based on the quasiclassical Green's function method
applicable to unconventional superconductor junctions \cite%
{Nazarov,Yokoyama}. We find that the resulting Josephson current in
D/DN/I/DN/D junctions is well fitted by the SR formula. Near the
transition temperature, it is proven analytically that Josephson
current follows the SR formula. We also confirm that this formula
does not hold in the ballistic junctions. It is clarified that in
the SR formula, the component of the pair potential which is
antisymmetric by the inversion operation around the plane
perpendicular to the interface is neglected. We also study $p$-wave
superconductor / diffusive normal metal /insulator/ diffusive normal
metal/ $p$-wave superconductor (P/DN/I/DN/P) junctions. %
%
The resulting Josephson current is also well fitted by $\cos \alpha
\cos \beta $, where $\alpha $ ($\beta$) denotes  the angle between the crystal axis of left (right) $p$-wave superconductor and the normal to the interface. This is a corresponding version of the SR formula in the $p$-wave
superconductor junctions. Furthermore, it is possible to extend the
theory for an unconventional superconductor (US) with arbitrary
angular momentum $l$. For US/DN/I/DN/US junctions, the expected
Josephson current is proportional to $\cos l\alpha \cos l\beta $.
The obtained results may serve as a guide for the analysis of the
experiments in unconventional superconductor junctions.

Before we proceed with a formal discussion, let us provide
qualitative arguments on the physical meaning of the SR formula and
explain why it holds in the diffusive junctions. First, we consider
$d$-wave superconductor junctions. The pair potentials of left and right $d$-wave superconductors are, respectively,  expressed
by $\Delta _{L}=\Delta \lbrack f_{SL}(\phi )+f_{ASL}(\phi )]\exp
(-i\Psi )$, and $\Delta _{R}=\Delta \lbrack f_{SR}(\phi
)+f_{ASR}(\phi )]$, with $f_{SL}(\phi )=\cos 2\phi \cos 2\alpha $ ,
$f_{ASL}(\phi )=\sin 2\phi \sin 2\alpha $, $f_{SR}(\phi )=\cos 2\phi
\cos 2\beta $, $f_{ASR}(\phi )=\sin 2\phi \sin 2\beta $ , where
$\phi $ is the injection angle measured from the interface normal,  $\Delta$ denotes the maximum value of the pair
potential and $\Psi$ is the phase difference across the junction. 
The terms proportional to $\cos
2\phi $, $i.e.$, $f_{SL}(\phi )$ and $f_{SR}(\phi )$, correspond to the $%
d_{x^{2}-y^{2}}$-wave pair potential and the the terms proportional to $\sin
(2\phi )$, $i.e.$, $f_{ASL}(\phi )$ and $f_{ASR}(\phi ) $, correspond to the
$d_{xy}$-wave pair potential, respectively. Here, $f_{SL}(\phi
)=f_{SL}(-\phi )$, $f_{SR}(\phi )=f_{SR}(-\phi )$, $f_{ASL}(\phi
)=-f_{ASL}(-\phi )$, and $f_{ASR}(\phi )=-f_{ASR}(-\phi )$ are satisfied. In
the actual calculation of Josephson current in D/DN/I/DN/D junctions, we
have to take an average over the various $\phi $. Due to the impurity
scattering in DN, the average is taken for the left and right D/DN interface
independently. Then we can drop $f_{ASL}(\phi )$ and $f_{ASR}(\phi )$ and
arrive at the SR formula, where only the terms $f_{SL}(\phi )$ and $%
f_{SR}(\phi )$ remain which do not change sign by exchanging $\phi $ for $%
-\phi $. This fact is in accordance with the recent result that the
proximity effect is absent in the case of $d_{xy}$-wave pair potential \cite{Nazarov}%
. Consequently, the resulting Josephson current is proportional to $\cos
2\alpha \cos 2\beta $.

%
%
%
Similar arguments apply to $p$-wave junctions. In this case, $f_{SL}(\phi
)=\cos \phi \cos \alpha $, $f_{ASL}(\phi )=\sin \phi \sin \alpha $, $%
f_{SR}(\phi )=\cos \phi \cos \beta $, and $f_{ASR}(\phi )=\sin \phi \sin
\beta $ are satisfied. The terms proportional to $\cos \phi $, $i.e.$, $%
f_{SL}(\phi )$ and $f_{SR}(\phi )$, correspond to the $p_{x}$-wave pair
potential and the the terms proportional to $\sin \phi $, $i.e.$, $%
f_{ASL}(\phi )$ and $f_{ASR}(\phi )$, correspond to the $p_{y}$-wave
pair potential. In the actual calculation for P/DN/I/DN/P junctions,
functions  $f_{ASL}(\phi )$ and $f_{ASR}(\phi )$ vanish after
averaging over angle $\phi $. This is consistent with our previous
results that the pair potential with $p_{y}$-wave symmetry does not
contribute to the proximity effect \cite{p-wave,Asano}.

\begin{figure}[htb]
\begin{center}
\scalebox{0.4}{
\includegraphics[width=19.0cm,clip]{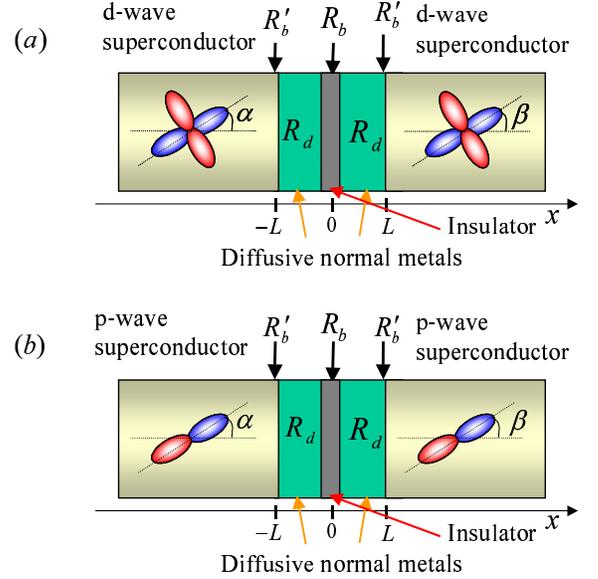}}
\end{center}
\caption{ (color online) Schematic illustration of the models of (a)
D/DN/I/DN/D and (b) P/DN/I/DN/P junctions.}
\label{f1}
\end{figure}

Next we formulate the junction model and basic equations starting from the $%
d $-wave case. We consider ballistic DID and D/DN/I/DN/D junctions.
The DN has a resistance $R_{d}$ and a length $L$ much larger than the mean
free path. The DN/D interfaces located at $x=\pm L$ have the resistance $%
R_{b}^{\prime }$, while the DN/I interface at $x=0$ has the
resistance $R_{b}$. We model infinitely narrow insulating barriers
by the delta function $U(x)=H^{\prime }\delta (x+L)+H\delta
(x)+H^{\prime }\delta
(x-L)$. The resulting transparencies of the junctions $T_{m}$ and $%
T_{m}^{\prime }$ are given by $T_{m}=4\cos ^{2}\phi /(4\cos ^{2}\phi +Z^{2})$
and $T_{m}^{\prime }=4\cos ^{2}\phi /(4\cos ^{2}\phi +{Z^{\prime }}^{2})$,
where $Z=2H/v_{F}$ and $Z^{\prime }=2H^{\prime }/v_{F}$ are dimensionless
constants and $v_{F}$ is Fermi velocity. Below we assume $Z \gg 1$. The schematic illustration
of the models  is shown in Fig. \ref{f1}. Here, $\alpha $ and $\beta $ denote the
angles between the normal to the interface and the crystal axis of the left
and right $d$-wave (or $p$-wave) superconductors, respectively. The lobe
direction of the pair potential and the direction of the crystal axis are
chosen to be the same. The pair potential along the quasiparticle trajectory
with the injection angle $\phi $ is given by $\Delta _{L}=\Delta \cos
[2(\phi -\alpha )]\exp (-i\Psi )$ and $\Delta _{R}=\Delta \cos [2(\phi
-\beta )]$ for the left and the right superconductor, respectively. For
ballistic junctions, we use a similar model without DN and calculate the
Josephson current following Ref. \cite{TK}.

We parameterize the quasiclassical Green's functions $G$ and $F$
with a function $\Phi _{\omega }$ \cite{Likharev,Golubov}:
\begin{equation}
G_{\omega }=\frac{\omega }{\sqrt{\omega ^{2}+\Phi _{\omega }\Phi _{-\omega
}^{\ast }}},F_{\omega }=\frac{{\Phi _{\omega }}}{\sqrt{\omega ^{2}+\Phi
_{\omega }\Phi _{-\omega }^{\ast }}}
\end{equation}%
where $\omega $ is the Matsubara frequency. Then the Usadel equation reads%
\cite{Usadel}
\begin{equation}
\xi ^{2}\frac{{\pi T_{C}}}{{\omega G_{\omega }}}\frac{\partial }{{\partial x}
}\left( {G_{\omega }^{2}\frac{\partial }{{\partial x}}\Phi _{\omega }}
\right) -\Phi _{\omega }=0
\end{equation}%
with the coherence length $\xi =\sqrt{D/2\pi T_{C}}$, the diffusion constant
$D$ and the transition temperature $T_{C}$. We solve the Usadel equation
with the boundary conditions in Ref.\cite{Yokoyama} at $x= \pm L$ and those
in Ref.\cite{Kupriyanov} at $x=0$.

The Josephson current is given by
\begin{equation}
\frac{{eIR}}{{\pi T_C }} = i\frac{{RTL}}{{2R_d T_C }}\sum\limits_\omega {%
\frac{{G_\omega ^2 }}{{\omega ^2 }}} \left( {\Phi _\omega \frac{\partial }{{%
\partial x}}\Phi _{ - \omega }^ * - \Phi _{ - \omega }^ * \frac{\partial }{{%
\partial x}}\Phi _\omega } \right)
\end{equation}
where $T$ is temperature and $R \equiv 2R_{d}+R_{b}+2R_{b}^{\prime}$ is the
normal state resistance of the junction. In the following we focus on the $%
I_C R$ value where $I_C$ denotes the magnitude of the maximum Josephson
current. We fix parameters as $Z^{\prime }=0$, $R_{d}/R_{b}=0.01$, $%
R_{d}/R_{b}^{\prime }=10$ and $E_{Th}/\Delta_{0}=1$ for D/DN/I/DN/D
junctions and $Z=10$ for DID junctions. $\Delta_{0}$ denotes the value of $%
\Delta$ at zero temperature. The choice of the small magnitude of
$Z^{\prime}$ and $ R_{b}^{\prime}$ and large Thouless energy is
justified by the fact that thin DN is naturally formed due to the
degradation of superconductivity near the interface.

\begin{figure}[tbh]
\begin{center}
\scalebox{0.35}{
\includegraphics[width=27.0cm,clip]{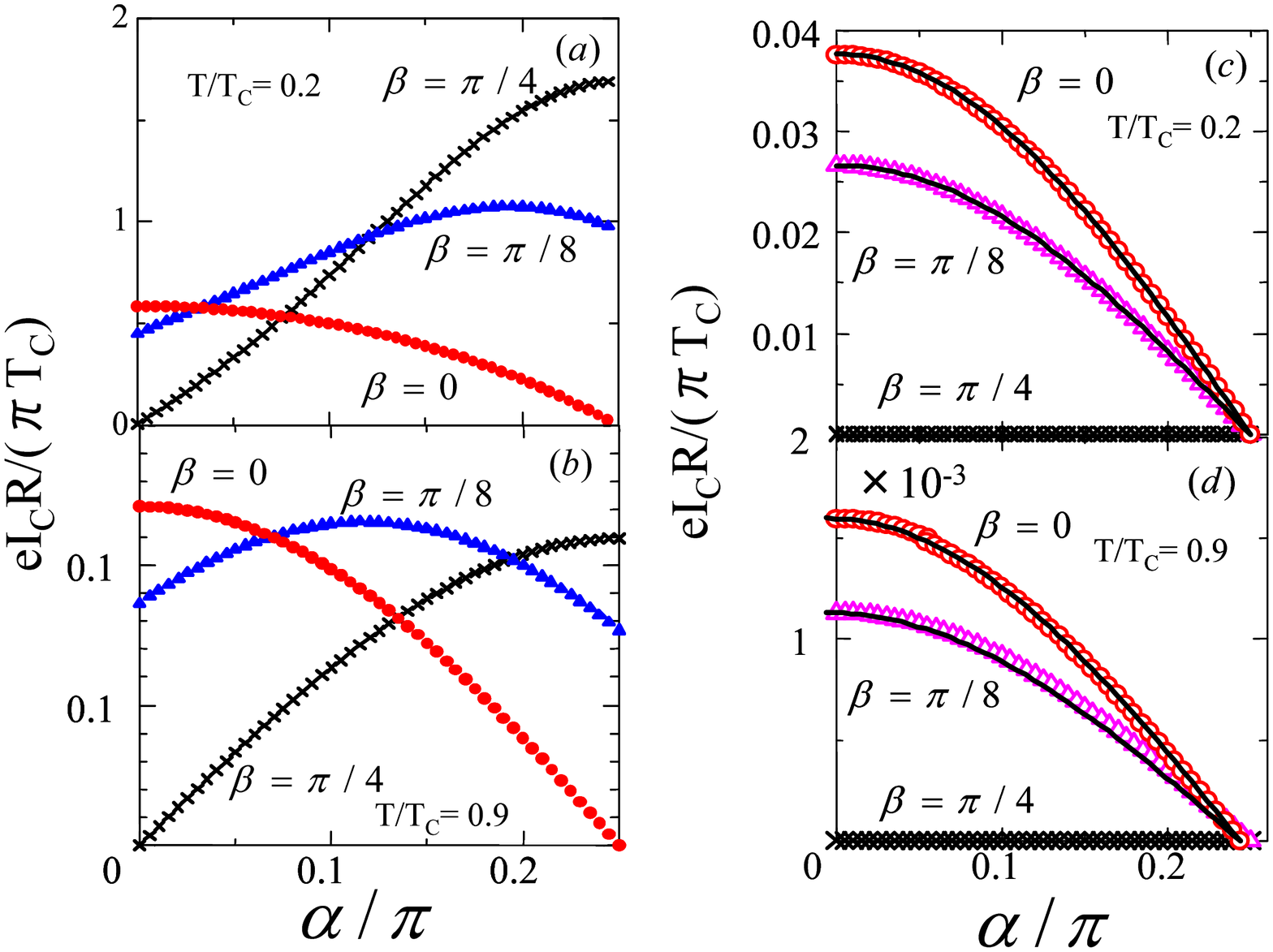}
}
\end{center}
\caption{(color online) Maximum Josephson current for $d$-wave
junctions. (a) and (b) DID junctions. (c) and (d) D/DN/I/DN/D
junctions. Solid lines in (c,d) are proportional to $\cos2\protect\alpha %
\cos2\protect\beta$. }
\label{f2}
\end{figure}

The $\alpha$ dependence of $I_{C}R$ for $d$-wave superconductor junctions is
plotted in Fig. ~\ref{f2}. In Figs.~\ref{f2}(a) and ~\ref{f2}(b), $I_{C}R$
of ballistic junctions is plotted for low ($T/T_{C}=0.2$) and high
temperature ($T/T_{C}=0.9$), respectively. With the increase of the
magnitude of $\beta$, the dependence of $I_{C}R$ on $\alpha$ transforms from
$\cos2\alpha$ to $\sin2\alpha$. These $\alpha$ dependences can not be
expressed by the SR formula, where $I_{C}R$ is proportional to $\cos2\alpha$
for fixed $\beta$. On the other hand in D/DN/I/DN/D junctions, $I_{C}R$ has
a simple form, $\cos2\alpha$, independent of $\beta$ at low and high
temperatures as shown in Figs.~\ref{f2}(c) and (d), respectively. The
magnitudes of $I_{C}R$ in D/DN/I/DN/D junctions are at least two orders
smaller than those in DID junctions. By taking account of the $\beta$
dependence, $I_{C}R$ is almost proportional to $\cos2\alpha\cos2\beta$. It
should be remarked that this fitting is possible for small magnitude of $%
Z^{\prime}$ and $R_{b}^{\prime}/R_{d}$ where the MARS formed at the D/DN
interface do not influence seriously the charge transport.

\begin{figure}[tbh]
\begin{center}
\scalebox{0.4}{
\includegraphics[width=22.0cm,clip]{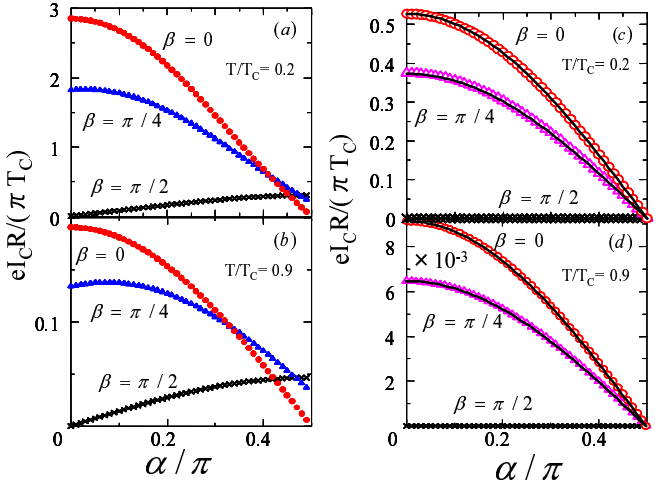}
}
\end{center}
\caption{(color online) Maximum Josephson current for $p$-wave junctions.(a) and (b) PIP junctions. (c) and (d) P/DN/I/DN/P
junctions with solid lines which are proportional to $\cos \protect\alpha %
\cos \protect\beta$. }
\label{f3}
\end{figure}

The corresponding results of $I_{C}R$ for $p$-wave superconductor
junctions are plotted in Fig. \ref{f3}. For P/DN/I/DN/P junctions,
$I_{C}R$ can be fitted by $\cos\alpha \cos\beta$. Similar to the
case of $d$-wave junctions,
this fitting is possible for small magnitude of $Z^{\prime}$ and $%
R_{b}^{\prime}/R_{d}$. For ballistic junctions, as shown in Figs.
\ref{f3}(a) and \ref{f3}(b), this fitting does not work any more.

In the following, we will present analytical result demonstrating why the SR
formula does not work in ballistic junctions and works in the diffusive
junctions. Although we focus on $d$-wave junctions, similar discussion is
possible for $p$-wave junctions. Near $T_C$ ($\Delta \ll \omega$), we can get
the formula for the ballistic DID junctions\cite{TK}:
\begin{equation}
\frac{eIR}{\pi T_C } = \frac{\Delta^{2} \sin\Psi} {8TT_C}F_{0},
\end{equation}
\begin{equation*}
F_{0}= <\cos ^2 2\phi > \cos 2\alpha \cos 2\beta + <\sin ^2 2\phi > \sin
2\alpha \sin2\beta.
\end{equation*}
Here, the average over the various angles of injected particles at the
interfaces is defined as
\begin{equation}
<B(\phi)> = \frac{\int_{-\pi/2}^{\pi/2} d\phi T(\phi)\cos\phi B(\phi)}{
\int_{-\pi/2}^{\pi/2} d\phi T(\phi)\cos\phi}  \label{aave}
\end{equation}
with $T(\phi)=T_{m}$. It is easy to check that $\left\langle {\cos ^2 2\phi }
\right\rangle$ and $\left\langle {\sin ^2 2\phi }\right\rangle$ are of the
same order for all $Z$. Therefore, SR formula cannot be applicable to
nonzero values of $\alpha$ and $\beta$. 
Also we can roughly estimate the Josephson current:
\begin{equation}
\frac{{eIR}}{{\pi T_C }} \cong \frac{{\Delta ^2 }}{{16TT_C }}\cos( 2\alpha-
2\beta) \sin \Psi
\end{equation}
which is consistent with the result in Fig. \ref{f2} (b). In the case of PIP
junctions, we can obtain the corresponding equation by replacing $2\alpha$, $%
2\beta$ and $2\phi$ with $\alpha$, $\beta$, and $\phi$ in the above
equations.

\begin{figure}[tbh]
\begin{center}
\scalebox{0.45}{
\includegraphics[width=15.0cm,clip]{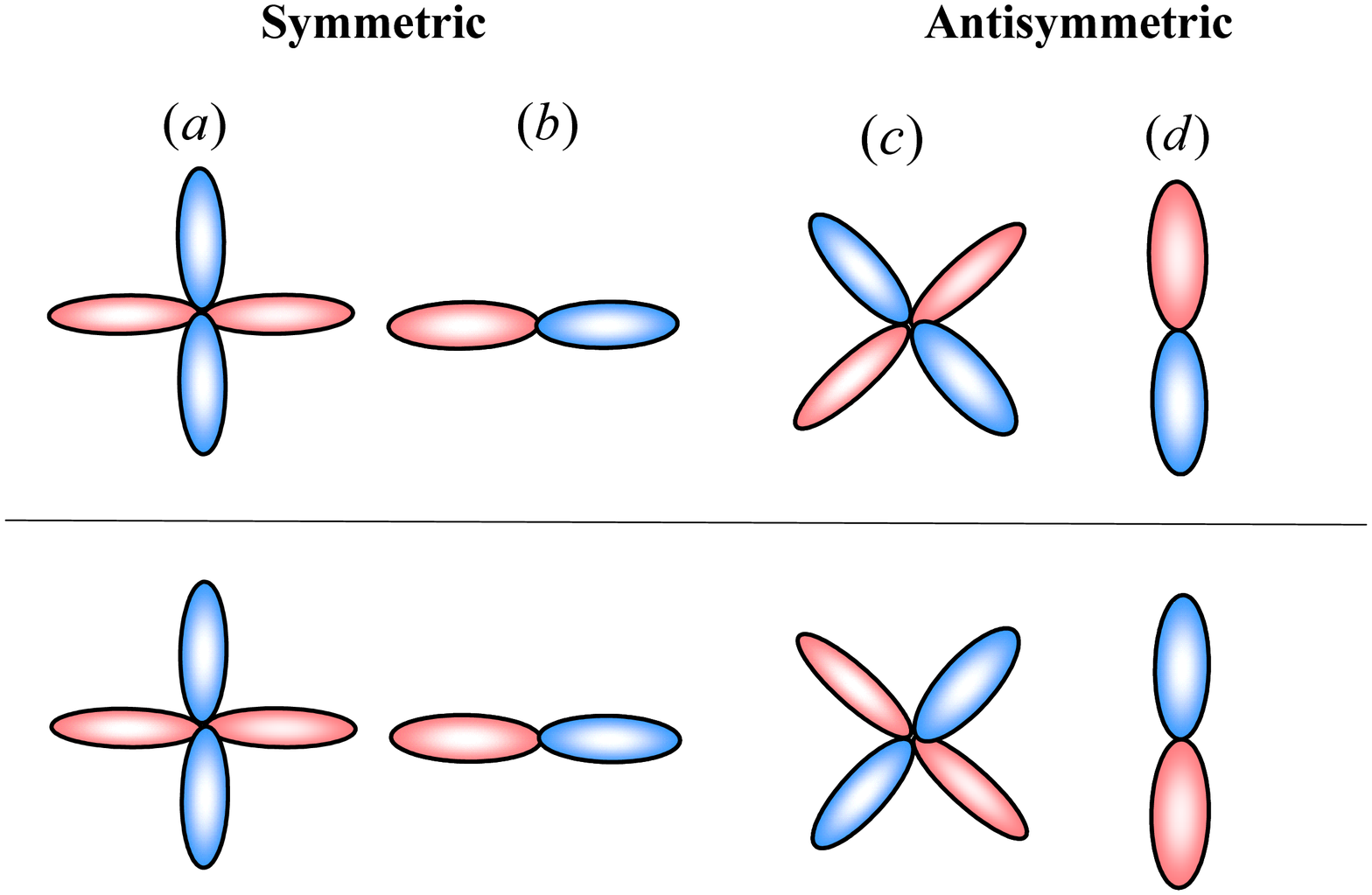}
}
\end{center}
\caption{(color online) Schematic illustration of the inversion symmetry of
the pair potential around the plane normal to the interface. (a)$%
d_{x^{2}-y^{2}}$-wave, (b)$p_{x}$-wave, (c)$d_{xy}$-wave,  and (d)$p_{y}$%
-wave. The $\protect\phi$ dependences are given by $\cos2\protect\phi$, $%
\sin2\protect\phi$, $\cos\protect\phi$, and $\sin\protect\phi$. }
\label{f4}
\end{figure}

Next we consider the D/DN/I/DN/D junctions. Near $T_C$, we can linearize the
Usadel equation as follows,
\begin{equation}
\xi ^2 \frac{{\partial ^2 }}{{\partial x^2 }}\Phi _{j \omega } - \frac{%
\omega }{{\pi T_C }}\Phi _{j \omega } = 0
\end{equation}
where j(=1, 2) denotes the left or right DN. Similarly the boundary
conditions at $x=-L$, $x=0$ and $x=L$ are reduced to
\begin{equation}
\frac{\partial }{{\partial x}}\Phi _{1\omega } \mid_{x=-L} = - \frac{R_d }{%
R^{\prime}_b L}\left( {\ - \Phi _{1\omega } + I_0 \cos 2\alpha e^{ - i\Psi }
} \right)\mid_{x=-L},
\end{equation}
\begin{equation}
\frac{\partial \Phi _{1\omega }}{{\partial x}}\mid_{x=0} = \frac{\partial
\Phi _{2\omega }}{{\partial x}}\mid_{x=0} = \frac{{R_d } \left( {\Phi
_{2\omega } - \Phi _{1\omega } } \right)} {{R_b L}} \mid_{x=0},
\end{equation}
\begin{equation}
\frac{\partial \Phi _{2\omega }}{{\partial x}}\mid_{x=L} = \frac{{R_d }}{{%
R^{\prime}_b L}}\left( {\ - \Phi _{2\omega } + I_0 \cos 2\beta }
\right)\mid_{x=L}
\end{equation}
with $I_0 =\Delta \left\langle {\cos 2\phi } \right\rangle $.

Solving the above equations, we find the expression for the Josephson current of the form
\begin{equation}
\frac{eIR}{\pi T_C} = \frac{R r r^{\prime 2} }{ R_d } \frac{T}{T_C }
\sum_\omega \frac{\gamma L <\cos 2\phi >^2 \Delta ^2 \cos 2\alpha \cos
2\beta \sin \Psi } {\omega^{2}F_{1}F_{2}},
\end{equation}
\begin{equation*}
F_{1}= \gamma L\sinh \gamma L + r^{\prime}\cosh \gamma L,
\end{equation*}
\begin{equation*}
F_{2}= [ (2rr^{\prime} + \gamma^2 L^2 ) \sinh \gamma L + ( 2r + r^{\prime})
\gamma L \cosh \gamma L ]
\end{equation*}
with $r = \frac{R_d }{R_b }$, $r^{\prime}= \frac{R_d}{R^{\prime}_b}$ and $%
\gamma = \sqrt{\frac{2\omega}{D}}$. Thus the SR formula is proven to be
valid near $T_C$. In the case of P/DN/I/DN//P junctions, $\cos 2\phi$, $
\cos2\alpha$ and $\cos2\beta$ have to  be replaced with $\cos \phi$, $\cos\alpha$ and $\cos\beta$, respectively to obtain the corresponding formula. This result is consistent with the previous
study of DID junctions with rough interface \cite{GK} where the SR formula
is applicable as well.

In order to understand the above results qualitatively, let's discuss the
symmetry of the pair potential by the inversion operation around the plane
perpendicular to the interface. As shown in Fig. 4, $d_{x^{2}-y^{2}}$-wave
and $p_{x}$-wave are symmetric while $d_{xy}$-wave and $p_{y}$-wave are
antisymmetric by this operation. Only the symmetric pair wave function is
taken into account in the SR formula.
Applying this idea to an arbitrary unconventional superconductor
with angular momentum $l$, one can argue that the Josephson current
is proportional to $\cos l\alpha \cos l\beta $. It is
straightforward to get this result just by replacing $\cos 2\phi $,
$\cos 2\alpha $ and $\cos 2\beta $ with $\cos l\phi $, $\cos l\alpha
$ and $\cos l\beta $ in Eq. (11), respectively.

In summary, we have studied the validity and the physical meaning of
the Sigrist-Rice formula in $d$-wave superconductor junctions.
According to the SR formula, the amplitude of the maximum Josephson
current is proportional to $\cos 2\alpha \cos 2\beta $.
 Although this formula is not applicable
to the ballistic junctions, it works well for D/DN/I/DN/D junctions
where the DN regions are located between
the $d$-wave superconductor and the insulator. We have also shown
that in P/DN/I/DN/P junctions, the Josephson current is proportional
to $\cos \alpha \cos \beta $. The obtained results may help to
obtain information about pairing symmetry in experiments with
unconventional superconducting junctions.

%
T. Y. acknowledges support by JSPS Research Fellowships for Young
Scientists. This work is supported by Grant-in-Aid for Scientific Research
on Priority Area "Novel Quantum Phenomena Specific to Anisotropic
Superconductivity" (Grant No. 17071007) from the Ministry of Education,
Culture, Sports, Science and Technology of Japan.

%



\begin{thebibliography}{99}
\bibitem{Tsuei} C.C. Tsuei and J.R. Kirtley, Rev. Mod. Phys. \textbf{72},
969 (2001).

\bibitem{Harlin} D.J. Van Harlingen, Rev. Mod. Phys. \textbf{67}, 515
(1995). %
%

\bibitem{SR95} M. Sigrist and T.M. Rice, Rev. Mod. Phys. \textbf{67}, 503
(1995).

\bibitem{Kashiwaya} S. Kashiwaya and Y. Tanaka, Rep. Prog. Phys. \textbf{63}
1641 (2000).

\bibitem{Sigrist2} M. Sigrist and T. M. Rice, J. Phys. Soc. Jpn. \textbf{61}%
, 4283 (1992).

\bibitem{Buch} L. J. Buchholtz and G. Zwicknagl, Phys. Rev. B \textbf{23},
5788 (1981); J. Hara and K. Nagai, Prog. Theor. Phys. \textbf{74}, 1237 (1986); C. Bruder, Phys. Rev. B \textbf{41}, 4017 (1990); C.R. Hu, Phys. Rev.
Lett. \textbf{72}, 1526 (1994).

\bibitem{Tanaka95} Y. Tanaka and S. Kashiwaya, Phys. Rev. Lett. \textbf{74},
3451 (1995).

\bibitem{TK97} Y. Tanaka and S. Kashiwaya, Phys. Rev. B \textbf{56}, 892
(1997).

\bibitem{TK} Y. Tanaka and S. Kashiwaya, Phys. Rev. B \textbf{53}, R11957
(1996).

\bibitem{Barash} Yu. S. Barash, H. Burkhardt, and D. Rainer, Phys. Rev.
Lett. \textbf{77}, 4070 (1996).

\bibitem{Testa} G. Testa, E. Sarnelli, A. Monaco, E. Esposito, M. Ejrnaes,
D.-J. Kang, S. H. Mennema, E. J. Tarte, and M. G. Blamire Phys. Rev. B
\textbf{71}, 134520 (2005).

\bibitem{Ilichev} E. Ilichev, M. Grajcar, R. Hlubina, R. P. J. IJsselsteijn,
H. E. Hoenig, H.-G. Meyer, A. Golubov, M. H. S. Amin, A. M. Zagoskin, A. N.
Omelyanchouk and M. Yu. Kuprianov, Phys. Rev. Lett. \textbf{86}, 5369 (2001).

\bibitem{Lombardi} F. Lombardi, F. Tafuri, F. Ricci, F. Miletto Granozio, A.
Barone, G. Testa, E. Sarnelli, J. R. Kirtley, and C. C. Tsuei, Phys. Rev.
Lett. \textbf{89}, 207001 (2002).

\bibitem{Bauch} T. Bauch, F. Lombardi, F. Tafuri, A. Barone, G. Rotoli, P.
Delsing, and T. Claeson, Phys. Rev. Lett. \textbf{94}, 087003 (2005).

\bibitem{Lee} Soon-Gul Lee and Yunseok Hwang, Appl. Phys. Lett. \textbf{76},
2755 (2000).

\bibitem{Harlingen} W.K.Neils, D.J. Van Harlingen, S. Oh, \textit{et. al.} ,
Physica C \textbf{368}, 261 (2002).

\bibitem{Smilde} H. J. H. Smilde, A. A. Golubov, Ariando, G. Rijnders, J. M.
Dekkers, S. Harkema, D. H. A. Blank, H. Rogalla, and H. Hilgenkamp, Phys.
Rev. Lett. 95, 257001 (2005).

\bibitem{Tsuei2} J. R.Kirtley, C. C. Tsuei, Ariando, \textit{et. al., }Nature
Physics \textbf{2}, 190 (2006).

\bibitem{Nazarov} Y. Tanaka, Y.V. Nazarov and S. Kashiwaya, Phys. Rev. Lett.
\textbf{90}, 167003 (2003); Y. Tanaka, Yu. V. Nazarov, A. A. Golubov, and S.
Kashiwaya, Phys. Rev. B \textbf{69}, 144519 (2004).

\bibitem{Yokoyama} T. Yokoyama, Y. Tanaka, A. A. Golubov, and Y. Asano, Phys.
Rev. B \textbf{73}, 140504(R) (2006).


\bibitem{p-wave} Y. Tanaka and S. Kashiwaya, Phys. Rev. B \textbf{70},
012507 (2004); Y. Tanaka, S. Kashiwaya and T. Yokoyama, Phys. Rev. B \textbf{%
\ 71}, 094513 (2005); Y. Tanaka, Y. Asano, A. A. Golubov and S. Kashiwaya,
Phys. Rev. B \textbf{72}, 140503(R) (2005).

\bibitem{Asano} Y. Asano, Y. Tanaka and S. Kashiwaya, Phys. Rev. Lett.
\textbf{96} 097007 (2006), Y. Asano, Y. Tanaka, T. Yokoyama and S.
Kashiwaya, Phys. Rev. B \textbf{74}, 064507 (2006).

\bibitem{Likharev} K.K. Likharev, Rev. Mod. Phys. \textbf{51}, 101 (1979).

\bibitem{Golubov} A. A. Golubov, M. Yu. Kupriyanov, and E. Il$^{\prime}$
ichev Rev. Mod. Phys. \textbf{76}, 411 (2004).



\bibitem{Usadel} K.D. Usadel, Phys. Rev. Lett. \textbf{25}, 507 (1970).

\bibitem{Kupriyanov} M. Yu. Kupriyanov and V. F. Lukichev, Zh. Exp. Teor.
Fiz. \textbf{94} (1988) 139 [Sov. Phys. JETP \textbf{67}, (1988) 1163].

\bibitem{GK} A. A. Golubov and M. Yu. Kupriyanov, JETP Lett. \textbf{67},
501 (1998); \textbf{69}, 262 (1999).
\end{thebibliography}
\end{document}